\documentclass[prd,aps,floats]{revtex4}

\usepackage[dvips]{graphicx}
\usepackage{amssymb}
\usepackage{amsmath}
\begin{document}

\title{The Cyclic Model Simplified\footnote{Talk given at Dark Matter 2004, Santa 
Monica, CA; February 18-20, 2004.}}

\author{ Paul J. Steinhardt$^{1,2}$
and Neil Turok$^{3}$}

\affiliation{$^1$Department of Physics, Princeton University,
  Princeton, New Jersey 08544, USA \\ $^2$ School of Natural Sciences,
  Institute for Advanced Study, Olden Lane, Princeton, New Jersey
  08540, USA \\
$^3$
Centre for Mathematical Sciences, Wilberforce Road, 
Cambridge CB3 0WA, U.K.
}

\date{April 2004}

\begin{abstract}
The Cyclic Model 
attempts to resolve the homogeneity, isotropy, and flatness problems 
and  generate a nearly scale-invariant spectrum of fluctuations during a period of 
slow 
contraction that precedes a bounce to an expanding phase.  
Here we describe at a conceptual level the
recent developments  that have greatly simplified our 
understanding of the 
contraction phase and the Cyclic Model  overall.  
The answers to many past questions and criticisms are 
now understood.
In particular, we show that the 
contraction phase has equation of state 
$w>1$ and that contraction with $w>1$ has a 
surprisingly similar properties to inflation with 
$w < -1/3$.  At one stroke, this 
shows how the model is different from inflation and 
why it may work just as well as inflation 
in resolving cosmological problems.   
\end{abstract}
\maketitle

\section{The Cyclic Manifesto}

	Two years ago, the Cyclic Model \cite{cyclic}
	was introduced as a radical alternative to the standard big bang/inflationary 
picture \cite{inflation}.
Its purpose is to offer  a new solution to the homogeneity, 
isotropy, flatness problems and a new 
mechanism for generating
a nearly scale-invariant spectrum of fluctuations. 

	One might ask why we should consider an alternative when inflation 
\cite{inflation,Bardeen,inflapert} has scored so many successes in explaining a 
wealth of new, 
highly precise data \cite{MAPspergel}.   There are several reasons.  First, 
seeking an 
alternative is just plain good science.  Science proceeds most rapidly when there 
are two or more 
competing ideas.  The ideas focus attention on what are the unresolved issues theorists must 
address and what are the important measurements experimentalists must perform.   
Inflation has had no serious competition for several years, and the result has 
been that its 
flaws have been ignored.  Many cosmologists are prepared to declare inflation to 
be 
established even though crucial experimental tests remain.  Competition stimulates critical thinking and removes complacency. 
 
	A second reason to consider an alternative is that, even though inflationary 
predictions are in marvelous accord with the data thus far, the theoretical front 
has seen 
little progress.  In fact, if anything, there has been retrogress.  The main 
questions about 
inflation that were  cited twenty years ago remain today.  What is the inflaton 
and why 
are its interactions finely-tuned?  How did the universe begin and why did it start to inflate?

	With the advent of string theory, these issues have become severe problems.  
Despite heroic efforts to construct stringy inflation models with tens or hundreds 
of 
moving parts (fluxes, branes and anti-branes) and examining a complex landscape of 
(at least) $10^{500}$ vacua, even a
single successful inflationary model is  difficult to construct
\cite{Malda}.  The notion that there is a landscape of $10^{500}$ or more string 
vacua 
has suggested to some that, if
there is  an acceptable vacuum somewhere, inflation 
 makes it possible 
 to populate all vacua; and that the ultimate 
 explanation for our universe is anthropic \cite{Susskind}.  
However, 
this cannot be the whole story since it begs the 
question of how the universe started in the first place.  No matter where you lie 
in the 
landscape, extrapolating back in time brings you to a cosmic singularity in a 
finite time.  
The issue of the beginning remains  unresolved. 

Furthermore, relying on the anthropic principle is like stepping on 
quicksand.  The power of a theory is measured by the ratio of its 
explanations/predictions to assumptions.  A good scientific theory
is observationally testable.  An anthropic explanation is based upon
considerations involving regions of space that are causally disconnected from us and that will, in many cases, never be observed by us.  
What parameters and properties can vary 
from region to region?  What is the probability distribution?  In models such 
as eternal inflation,  the relative likelihood of our being in one region or another is ill-defined since there is no unique time slicing and, therefore, 
no unique way of assessing the number of regions or their volumes.
Brave souls have begun to head down this path, but
it seems likely to us to drag a beautiful science towards the darkest
depths of metaphysics.

Another unresolved issue is trans-Planckian effects on the production
of density perturbations \cite{trans}.  
In inflationary cosmology, the fluctuations observed in the cosmic
microwave background had wavelengths at the beginning of inflation that 
were smaller than the Planck scale.  
The standard approximation is to assume the initial distribution of
sub-horizon and, hence, sub-Planckian fluctuations 
corresponds to quantum fluctuations on an empty, Minkowski
background.  However, quantum gravity effects may cause the 
distribution to be different on sub-Planckian wavelengths.  The 
unknown distortion would be inflated and produce an uncertain 
correction to inflationary predictions for the cosmic microwave background
anisotropy.

	Finally, the big bang/inflationary picture is still reeling from the recent shock 
that 
most of the universe consists of dark energy \cite{SN}.  The concept had been 
that, once conditions 
are set in the early universe, the rest of cosmic evolution is simple.  Dark 
energy has 
shattered that dream.  Dark energy was not anticipated and plays no significant 
role in the 
theory.  Observations have forced us to add dark energy {\it ad hoc.}

	The current approach in big bang/inflationary model-building has been to 
treat the 
key issues --  the bang, the creation of homogeneity and density fluctuations, and 
dark 
energy --  in a modular way.  
Separate solutions with separate ingredients are sought for 
each.  Perhaps this approach
will work, all the problems cited above will be resolved, and a simple 
picture will emerge.  But, perhaps the time has come to consider a different, 
holistic 
approach.

	The cyclic model has an ambitious manifesto.
Its goal is  to address the entire history of the 
universe, past and future, in an efficient, unified approach.  There is one 
essential 
ingredient -- branes in the higher-dimensional picture or a scalar field in the 
four-dimensional effective theory 
--  that is simultaneously responsible for explaining the big 
bang; the solution to the homogeneity, isotropy, flatness,  and monopole
problems \cite{cyclic}; the 
generation of nearly scale-invariant fluctuations that seed large-scale structure 
\cite{ekperts,newtolley};  and, the source of dark 
energy \cite{cyclic}.  Simplicity and parsimony are essential elements.  The range 
of acceptable parameters is broad \cite{design}.

	Over the past two years, the Cyclic Model has progressed  remarkably.  The 
concept has been examined by numerous groups, and many, many useful criticisms and 
questions have been raised \cite{lyth,linde,contrascale,durrer,ambig,string}.    
As we and our collaborators have tried to 
address these issues, the results have been interesting.  
First, we have discovered that the Cyclic Model already contained 
the answers.  Not a single new ingredient has had to be added thus far.  Rather, 
we have 
learned to recognize fully the physical properties of the components the model  
contained at the 
outset \cite{reply,bcbb,duality,dual2,chaos}.    That is, we have been
discovering new physical principles stemming from the original model 
rather than 
 adding new ingredients and patches.  Second, as we have come to 
understand the Cyclic Model better, the picture has become much, much simpler.  We 
believe we can stick by our manifesto:
If the 
model is going to work, it will be because of basic ideas as simple 
and compelling as inflation.  In fact, we find that there are remarkable, 
unanticipated  
parallels between inflationary expansion
and the contracting and bounce phases of the Cyclic Model \cite{duality,dual2}.  
There remain important open issues about the bounce itself, but, now we can 
confidently 
say that many of the issues that plagued previous attempts at contracting 
cosmological 
models have been cleared away and there are solid reasons for optimism about 
resolving 
the remaining issues.

The purpose of this essay is to present the simplified view of the Cyclic 
Model, 
focusing on the stages that are most novel and controversial: the contraction and 
bounce. 
We focus on the two key ingredients needed to understand 
the contracting phase: branes and the 
equation of state $w>1$.  As we explain, the two features lead to a 
series of novel
physical effects that  solve the 
homogeneity, isotropy, and flatness problems 
and
ensure a nearly scale-invariant spectrum of density perturbations following the 
big bang.  

\section{The basic concept}

The Cyclic Model was developed  based on the three intuitive notions:
\begin{itemize}
\item the big bang is not a  beginning of time, but rather a transition to an 
earlier phase of 
evolution; 
\item the evolution of the universe is cyclic;
\item the key events that shaped the large scale structure of the universe 
occurred during 
a phase of slow contraction before the bang, rather than a period of rapid 
expansion 
(inflation) after the bang.
\end{itemize}

The last point means that, unlike previous periodic models, the 
cycles are tightly interlinked.  The events that occurred a cycle ago
shape our universe today, and the events occurring today will shape our
universe a cycle from now.  It is this aspect that transforms the
metaphysical notion of cycles into a scientifically testable concept.  We can make 
physical measurements today that determine whether the
large scale structure of the universe was set before or after the bang.

The model is motivated by the M-theoretic notion
that our universe consists 
of two branes separated by a microscopic gap (the ``bulk'') \cite{mtheory}.  
Observable particles -- quarks, 
leptons, photons, neutrinos, etc. --  lie on one brane and are constrained to move 
along it.   
Any particles lying on the other brane can interact gravitationally with particles 
on our 
brane, but not through strong or electroweak interactions.   So, from our 
perspective, particles on the other brane
are a dark form of 
 matter that cannot be detected in laboratories looking for weakly interacting 
particles. (The Cyclic Model does not predict whether most of the 
dark matter detected cosmologically is weakly interacting particles
on our brane or particles lying on the other brane.  Both are 
logical possibilities.)

In an exactly supersymmetric vacuum state, the branes do not interact at all.  The 
virtual 
exchanges of strings and membranes  cancel so that there is no force 
attracting or 
repelling them.  We conjecture that, in a realistic (supersymmetry
breaking) vacuum state, an attractive, 
spring-like 
force does attract them \cite{cyclic}. Specifically, we imagine that the 
force is very weak when 
the branes are thousands of Planck distances apart (as they would be now), so that 
they 
are hardly moving. However,
the force increases in strength as the branes draw together.  
Equivalently, we assume an interbrane potential of the form shown in Figure 1, 
where 
here $\phi$ is the moduli field that determines the interbrane separation.  When 
the 
branes are far apart, the potential is flat and nearly positive; as the branes 
draw together, 
the potential falls steeply and becomes negative.  When the branes come within a 
string-scale distance  
 apart (corresponding to $V \approx V_{end}$ in the Figure), 
the potential disappears exponentially.  Collision corresponds to $\phi 
\rightarrow -\infty$.  
The scenario can be described 
by an effective four-dimensional theory for $\phi$, where 
$\phi$ runs back and forth the 
potential from some positive value (corresponding to the present brane separation) 
to $-
\infty$ and back.

\begin{figure} 
\begin{center}
\includegraphics[clip=true,scale=0.7]{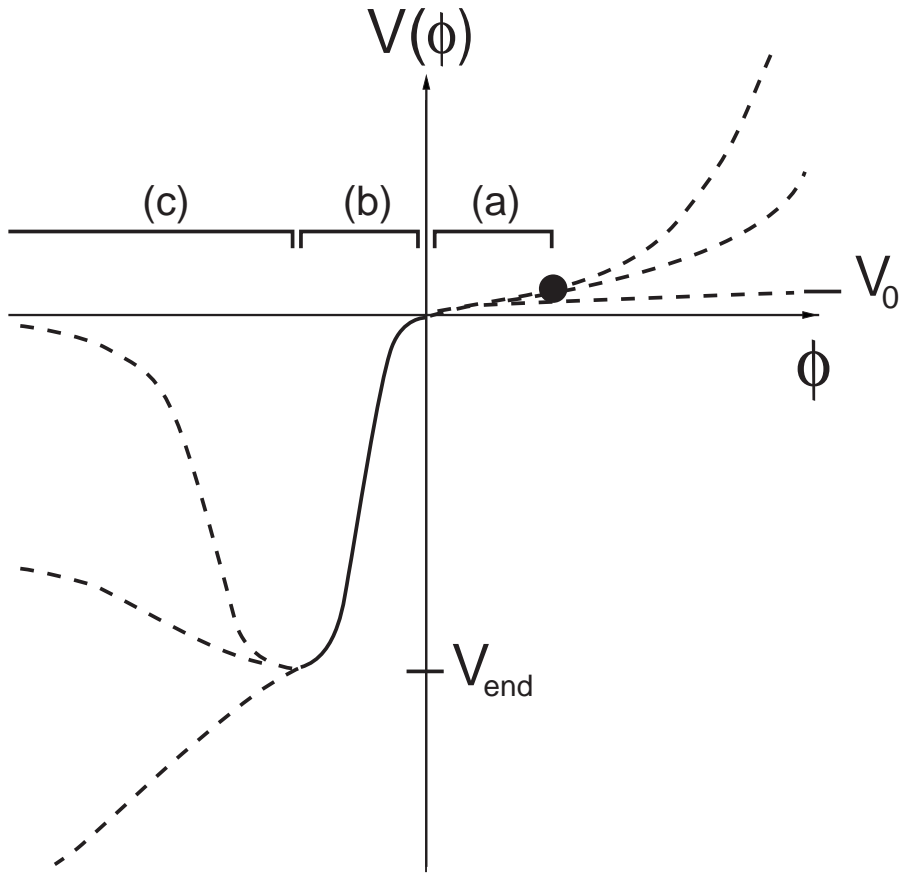}
\end{center}
\bigskip
\caption{Scalar potentials suitable for a cyclic universe model.
Running forward in cosmic time, Region (a) governs the decay of
the vacuum energy, leading to the end of the slow acceleration
epoch.
Region (b) is the region where scale invariant perturbations
are generated. In Region (c), as one approaches the big crunch
($\varphi \rightarrow -\infty$), the kinetic energy
 dominates.}
 \label{pot}
 \end{figure}

The interbrane potential
causes the branes to collide at regular intervals.  The collision itself
is the big 
bang.  The bang
is slightly inelastic, infusing the universe with new matter and 
radiation.  From the four-dimensional effective theory, the kinetic
energy of $\phi$ is dominant for a 
brief period after the bounce, but it decreases rapidly as the 
universe expands.
Hence, after the branes bounce 
apart, the branes slow down to essentially a halt, and the universe becomes 
radiation-
and matter-dominated.  The heat from the collision dominates the universe for a 
few 
billion years, but eventually it is diluted enough that the positive interbrane 
potential 
energy density dominates.  This acts as a source of dark energy that causes the 
expansion 
of the branes to accelerate.  The matter, radiation, and large scale structure are 
all diluted away 
exponentially over the next trillion years or so, and the branes become 
nearly 
perfect vacuua.  However, the interbrane attractive force ensures that the 
acceleration only 
lasts a finite time.  Inexorably, the branes are drawn together and the potential 
energy 
decreases from positive to negative values. The acceleration stops and, once the 
potential 
decreases to the point where $V= -\frac{1}{2}\dot{\phi}^2$, the total energy 
density is zero and the Hubble expansion rate becomes zero.  The universe 
switches 
from expansion to contraction.   The branes themselves do not contract 
or 
stretch significantly. Rather, the distance between them shrinks as the two branes 
crash together.  That is, the contraction only occurs in the extra dimension 
between the 
branes. The collision is a singularity in the sense that a dimension momentarily 
disappears.  However, the branes exist before, during and after the collision, 
which plays a crucial role in tracking what happens to the universe
through the bounce.

During the dark energy dominated phase, the branes are stretched to the point 
where they 
are flat and parallel.  During the contraction phase, the branes stop stretching and quantum 
fluctuations naturally cause the branes to wrinkle.  
Due to the  wrinkles,
 the branes do 
not collide everywhere at the same time.  Since the collision creates matter and 
radiation, 
this means that different regions heat and expand at different times.  The result 
is that the 
universe is slightly inhomogeneous after the collision. For an exponentially steep 
interbrane potential, the spectrum of temperature fluctuations is nearly scale-
invariant \cite{cyclic,ekperts,newtolley}.

Unlike cyclic models discussed in the 1920s and 30s, the entropy density does not 
build 
up from cycle to cycle.  Here is an example of where we take full advantage of the 
idea of 
branes and extra dimensions:  The entropy created in one cycle is expanded and 
diluted to near zero density after the dark energy dominated phase, but the
entropy density  does not 
increase again in the contraction phase. The simple reason is
that the branes themselves 
do not contract.  Only the extra dimensions contract.  

From a local observer's point of 
view, the entropy density undergoes precise cyclic behavior.  Yet, the total 
entropy on the 
branes grows, in accord with the second law of thermodynamics.  It is just 
that entropy 
is being exponentially diluted from one cycle to the next, so any
given local observer cannot detect the  entropy 
from previous cycles.  

The collisions can continue indefinitely despite the fact that the brane 
collisions are 
inelastic because gravity supplies extra energy during each contraction phase.  
During 
contraction, the kinetic energy of particles or, in this case, branes, is blue 
shifted due to 
gravity.  This simply means gravity is providing extra kinetic energy 
in addition to  what the 
interbrane force produces.  So, when the branes collide, it is with greater energy 
than 
would be obtained with the interbrane force alone.  The net result is that gravity 
adds to 
the kinetic energy which converts partially to matter and radiation.  A key result 
(shown 
in Ref. \cite{cyclic}) is that, if we consider the coupled gravity, scalar field, 
and radiation 
evolution equations, there exists a cyclic solution that is stable under small 
perturbations.

\section{Parsimony: An efficient use of space-time}

The Cyclic Model is more parsimonious than
inflation in that a greater proportion of space-time looks like the universe
we see.  In inflationary models, most of space-time consists either
of a very high energy inflating phase, or of the empty vacuum 
to which bubble interiors tend at late times.
With exponential rarity, bubbles are formed in the
high energy phase, and, within each, a hot big bang 
bang phase forms. 
The interior of the bubble is hot at first, but the temperature
and density decrease steadily with time, and structure formation
stops once dark energy dominates the universe.
Hence, along any time-like world-line in the inflating universe,
there is only a single brief
interval (when  the world-line crosses a bubble wall)
in which there exist stars and galaxies.
In the cyclic model, every world-line has repeated, periodically spaced
intervals in which stars and galaxies form.

The description of inflation above made the conventional
assumption that the interior of
a bubble never undergoes further high-energy inflation.  If the
dark energy is due to a cosmological constant, though, this may
not be the case.  Imagine a quadratic inflaton potential, say, whose
minimum has a small, positive value corresponding
to the currently observed dark energy density.  High-energy
inflation occurs when the inflaton field lies far from the minimum,
high up the potential.  Inflation ends in a region when the field
falls to the minimum. This region is equivalent to a bubble.
However, here the minimum corresponds to a
low-energy de Sitter phase.  With infinitesimally small
probability, de Sitter
fluctuations can carry the inflaton field  back up
up the potential high enough to begin a second period of high-energy
inflation followed by a second bubble and big bang phase.  In this
case, a time-like  world-line would have irregularly spaced intervals in which
stars and galaxies form.  However, even in this case,
the intervals would be exponentially
far apart compared to the model  with periodic cycling.

By either reckoning, inflation wastes space-time.
In a Bayesian
comparison of the two theories, more wasted space-time translates into a 
reduced
probability of a theory being correct.
If $P(A)$ is the probability of theory $A$ and if $P(O|A)$ is the
probability of observation $O$ given theory $A$, then
\begin{equation}
\frac{ P({\rm inflation})}{P({\rm cyclic})} =
\frac{P({\rm stars} \, | \, {\rm inflation})}
{P({\rm stars} \, | \, {\rm cyclic})}  \ll 1,
\end{equation}
assuming equal priors for the two theories.
(A similar analysis is sometimes used to explain why
inflation is more desirable than the standard big bang model.)
We make this point for amusement purposes only.  At this point in time,
it seems plausible to assign the models equal priors.  However,
we hope that
future observations and
developments in fundamental physics will be the decisive factors.

\section{Forever cycling?}

The  description in the previous section  is an idealization, because there is dissipation from cycle to cycle \cite{nobel}.  
For example, black holes formed during one cycle will
survive into the next cycle, acting as defects in an otherwise nearly
uniform universe. (In the vicinity of the black holes, there is no cycling
due to their strong gravitational field.)   Also, quantum
fluctuations and thermal fluctuations will, with exponentially small rarity, create
'bad regions' which fall out of phase with the average cycling and could
form giant black holes \cite{bubble}.  
In comoving coordinates, the black holes
and bad regions increase in density over time.
In this sense, the comoving observer sees the universe as ``winding down." 
Similarly, a local observer will see the cycling as having finite duration  in the
sense that, at some point, after many, many cycles, he will end up inside a
black hole (or bad region) and cease to cycle.  Thus, we conclude that cycling conserves
energy and is not perfectly efficient; it is neither perpetual motion of the first or 
second kind.  
However, because of the stretching of space, the distance between the
defective regions remains larger than Hubble distance.  New cycling regions
of space are being created although any one region of space
cycles for a finite time.  The cyclic model thereby satisfies the 
conventional thermodynamic laws even though the cycling continues forever.

It has been suggested that the holographic principle may place a stronger 
constraint on the duration of cycling \cite{susskind}.  
The  argument is based on the 
fact that there is an average positive
energy density per cycle.  Averaging over many cycles, the cosmology can be viewed
  as an expanding de Sitter Universe. A de Sitter universe has a finite horizon with a maximal entropy 
  within any observer's causal patch \cite{Lenn2}
  given by the surface area of the horizon. 
  Each bounce produces a finite entropy density or, equivalently,
  a finite total entropy within an observer's horizon. Hence,
the maximal entropy is reached after a 
finite number of bounces.  
(Quantitatively, a total entropy of $10^{90}$ is produced within an 
observer's horizon
each cycle, and the maximum entropy within the horizon is $10^{120}$, leading to a limit of $10^{30}$ bounces.)

\begin{figure} 
\begin{center}
\includegraphics[clip=true,scale=0.7]{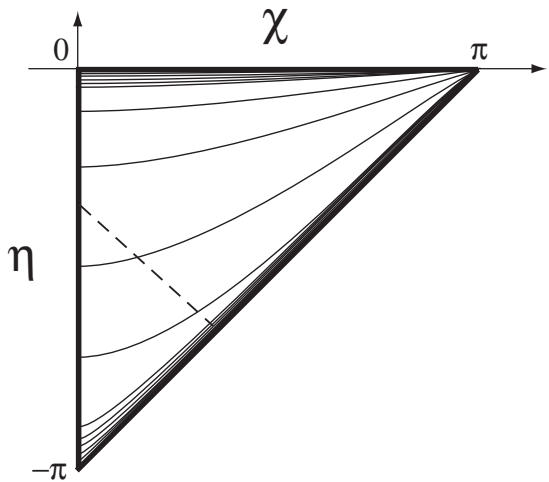}
\end{center}
\bigskip
\caption{The cyclic model has an average positive energy 
density per cycle, so its conformal diagram is similar
to an expanding de Sitter space with constant density.  
The bounces occur along flat slices
(curves) that, in this diagram, 
pile up near the diagonal and upper boundaries.   
For true de Sitter space, entropy bounds limit the total entropy 
in the entire spacetime.  For the cyclic model, the bounds
only limit the entropy between caustics (the bounces). 
Particles or light-signals emitted 
in an earlier cycle (or before cycling commences)
are likely to be scattered or annihilated as they
 travel through many intervening cycles
 (dashed line) to reach a present-day observer.    
The observer is effectively insulated from what preceded
the cycling phase, and there are no measurements to determine 
how many cycles have taken place.
 }
 \label{conf}
 \end{figure}

Closer examination reveals a flaw in this analysis \cite{nobel,kleban}.
Although the overall causal structure of the four-dimensional effective theory may be de Sitter, it is punctuated by bounces in which 
the scale factor approaches zero. See Figure~\ref{conf}.
Each bounce corresponds to a spatially flat caustic surface.  All known entropy bounds used in the holographic principle do not apply to surfaces which cross caustics. Hence, holographic bounds can be found for regions of space between a pair of caustics ({\it i.e.}, within a single cycle), but there is no surface extending across two or more bounces for which a valid entropy bound applies. If the singular bounce is replaced by 
a non-singular bounce at a small but finite value of the scale factor, the same conclusion holds. In order for a contracting universe to bounce at a finite value of the scale factor, the null energy condition must be violated.  However, 
the known  entropy bounds  require that 
the null energy condition be satisfied. 
Once again, we conclude that the 
entropy bounds cannot be extended across more than one cycle.  
Yet another way of approaching the issue is to note that
both singular and non-singular bounces have the property that light rays focusing during the contracting phase defocus after the bounce, which violates a key condition required for entropy bounds. In particular, the light-sheet
construction used in covariant entropy bounds \cite{bousso} are restricted to  surfaces that are uniformly contracting, whereas the extension of a contracting light-sheet across a bounce turns into a volume with expanding area. 
Hence, if bounces are physically possible,  entropy bounds do
not place any restrictions on the number of bounces.

Does this mean that the cycling has no beginning?  
This issue is not settled at present \cite{grat1}.
We have noted that the cyclic model has the causal structure of an expanding 
de Sitter space
with bounces occurring on flat spatial slices.
For de Sitter space, the expanding phase is geodesically incomplete, so 
the cyclic picture cannot be the whole story.  The most likely story is
that cycling was preceded by some singular beginning.
Consider a universe that settles into  cycling  beginning 
from some flat slice in the distant past many bounces ago.
Any particles produced before cycling must
travel through an 
exponentially large  number of bounces, each of which is a caustic surface with a 
high density of matter and radiation at rest with respect 
to the flat spatial slices.  
Any particle attempting this trip 
will be scattered or  annihilated and its information will be thermalized
before reaching a present-day observer.   
Consequently, the observer is effectively insulated from what preceded
the cycling phase, and there are no measurements that can be made to determine 
how many cycles have taken place.
Even though the space is formally
geodesically incomplete, it is as if, for all practical purposes, the universe
has been cycling forever.
We are currently exploring if this picture can be formalized.

\section{Contraction and Bounce}

Major progress has been made in understanding the  most controversial stages of 
the 
Cyclic Model:  the contraction and bounce.  Concerns about these stages are 
understandable.   Previous attempts to construct cyclic or oscillatory models all 
failed 
due to various problems that arise during a contraction phase:  the matter and 
radiation 
density diverge; the entropy density diverges; the 4-curvature diverges;  the 
anisotropy, 
spatial curvature, and inhomogeneity diverge; collapse exhibits chaotic mixmaster 
behavior.   Hitherto, this pathological behavior has rendered it inconceivable that a nearly homogeneous, isotropic and flat universe with small-amplitude  scale-invariant fluctuations could emerge from a bounce.

We now understand that the Cyclic Model can evade these problems because of two 
distinctive properties:

\noindent
$(i)$  {\it Since matter and radiation are confined to branes, their background densities do  not diverge at the bounce.  New entropy is created but old entropy remains dilute \cite{cyclic}.} Unlike previous cyclic models, the entropy density 
does not 
build up from cycle to cycle.  Instead, the entropy density returns to near zero 
towards the end of each cycle. 

\noindent
$(ii)$
{\it Because $w \gg 1$ during the contraction phase, the universe is homogeneous, 
isotropic 
and flat \cite{chaos} with  a scale-invariant spectrum of density perturbations 
\cite{duality,dual2}.  The $w \gg 1$ condition 
also ensures that anisotropies are small and first order perturbation theory remains valid until just 
before the bounce \cite{chaos}.}  

\vspace{.05in}

These effects due to branes and a $w>1$ energy component 
are novel and critical to the success of a cyclic scenario.
 Earlier attempts at cyclic models over the last 
century did not include branes because
that concept came into vogue only during the 
last decade.  However, one might naturally wonder why $w>1$ was not considered 
previously.  The probable reason is that, prior to inflation, cosmologists often 
assumed 
for simplicity that the universe is composed of ``perfect fluids'' 
for which $w= c_s^2$,
where the equation of state  $w$ equals the ratio of pressure $p$ to energy 
density $\rho$, 
and the speed of sound $c_s$ is defined by $c_s^2 = d p/d\rho$.  If $w>1$ and the 
fluid 
is perfect, then $c_s >1$, which is physically disallowed for any known fluid.  
With the 
advent of inflation, cosmologists have become
more sophisticated and flexible about what 
fluids they are willing to consider.  The inflaton, for example, has $w\approx -
1$, yet the 
speed of sound is positive and well-behaved.  A rolling scalar field with 
canonical kinetic 
energy has $c_s=1$.  Similarly, it is possible to have $w>1$ and yet $0 \le c_s^2  
\le 1$ 
without violating any known laws of physics.  This opens the door to
a novel kind of cyclic model.

\subsection{What is the $w>1$ component?}
\label{sec31}

A $w>1$ energy component did not have to be added to the Cyclic Model
in order resolve the traditional problems of contracting universes.
The component  was there 
from the very beginning waiting 
for its effects to be recognized \cite{cyclic,duality}.    The $w > 1$ equation of 
state is directly due to the interbrane potential that draws the branes together.
 The interbrane separation is described by a modulus
field $\phi$ with an 
attractive potential that is positive when the branes are far apart and becomes 
negative as 
the branes approach.  The rolling from a positive to a negative value is necessary 
for 
switching the universe from accelerated expansion to contraction.   To see how 
this 
occurs, consider the Hubble parameter after 
the universe is dominated by the scalar field and its potential:
\begin{equation}
H^2 = \frac{8 \pi G}{3}[\frac{1}{2} \dot{\phi}^2 +V].
\end{equation}
The universe is spatially flat after a period of accelerated 
expansion, so we have included only the scalar field kinetic and 
potential energy density terms.  
In order to reverse from expansion to contraction, there must be some time 
when $H$ hits zero.  Since the scalar field kinetic energy density is positive 
definite, the 
only way $H$ can be zero is if $V<0$.  So, reversal from accelerated
expansion 
forces us  to have $V$ roll from a 
positive value (where $V$ as the dark energy) to a negative value.
 However, this automatically creates an equation of state
\begin{equation}
w \equiv \frac{ \frac{1}{2} \dot{\phi}^2 -  V} { \frac{1}{2} \dot{\phi}^2 + V} > 1
\end{equation}
when $V<0$ (during the contraction phase).  If the potential is exponentially 
steep, 
$\propto -e^{- c \phi}$ with $c>>1$, it is straightforward to show that it has an 
attractor 
solution with equation of state $w=(c^2 -3)/3  \gg  1$. 
Hence, a $w>1$ component is an essential, built-in feature of the 
Cyclic Model.

\subsection{$w>1$ and the homogeneity, isotropy and flatness problems}

We first consider the effect of a $w>1$ energy 
component  on the average homogeneity, isotropy and 
flatness of the universe.   

As a warm-up, it is useful to recall  how inflation homogenizes, isotropizes and 
flattens 
the universe. In the standard big bang/inflation model, we  imagine that the 
universe 
emerges from the big bang with many elements contributing to the Friedmann 
equation:
\begin{equation}
H^2 = \frac{8 \pi G}{3}[\frac{\rho_m}{a^3}+ \frac{\rho_r}{a^4} + 
\frac{\sigma^2}{a^6}
+ \ldots + \rho_I] - \frac{k}{a^2},
\end{equation}
where $H$ is the Hubble parameter; $a$ is the scale factor; $\rho_{m,r}$ is the 
matter 
and radiation density; $\sigma^2$ measures the anisotropy; $k$ is the spatial 
curvature 
and $\rho_I$ is the energy density associated with the inflaton.  The parameters 
$\rho_i$ 
and $\sigma$ are 
constants which characterize the condition when $a=1$, which we can 
choose without loss of generality to be the beginning of inflation.  Each energy 
density 
term decreases  as   $1/a^{3(1+w)}$.
 For the 
inflaton, we have assumed $w\approx -1$ and the energy density is nearly 
$a$-independent. The  ``$\ldots$'' refers to other possible energy components, 
such as the 
energy associated with inhomogeneous, spatially varying fields.   Inflation works 
because, as the universe  expands,
all other contributions, including 
the spatial curvature and anisotropy,  are shrinking quickly as $a$ grows, while 
the inflaton density $\rho_I$ is nearly constant.  
Once the 
inflaton dominates, the future evolution is determined by its behavior and its 
decay 
products.  A complex initial state is focused into a very simple 
condition after sufficient 
expansion, a spacetime that is  homogeneous, isotropic and spatially flat.

Now let's consider the same equation in a contracting universe.  
In this case, we do not  
need an inflaton, but one might imagine a finite cosmological constant instead.  
In a 
contracting phase, the cosmological constant becomes rapidly unimportant as the 
universe contracts and $a$ shrinks.  (Hence, 
an expanding de Sitter phase is stable to small 
perturbations, but a contracting one is not.)  The term that will naturally 
dominate is the 
one that grows the fastest as $a$ shrinks.  In this case, the anisotropy term, 
$\sigma^2/a^6$ wins out.   A more careful analysis including the full Einstein 
equation 
 reveals that the universe not only becomes anisotropic, 
but also develops a large 
anisotropic spatial 
curvature and enters a phase of chaotic mixmaster behavior as the 
bounce 
approaches.  The anisotropy causes the universe to stretch in one direction while 
collapsing in other directions, forcing the geometry to become cigar-like.  Next, 
the anisotropic 
spatial curvature grows rapidly, forcing the anisotropic stretching to stop and 
switch 
direction.  When the anisotropic stretching makes the universe cigar-shaped again, 
the 
spatial curvature term once again causes the anisotropic stretching and 
another switch in  direction.  
The evolution follows a chaotic path in which the bounces from one stretch 
direction to 
another appears to follow a random pattern.  This behavior is called ``chaotic 
mixmaster 
behavior'' \cite{BKL}.  The mixmaster behavior produces severe inhomogeneities.  
The anisotropic contraction follows a set of dynamical equations with chaotic 
solutions.  
Hence, the directions of stretching and contracting seem to follow a random 
pattern that 
is exponentially sensitive to initial conditions. As a result, if the universe is 
even slightly 
inhomogeneous when the contraction phase begins, it will become increasingly 
inhomogeneous as different 
regions undergo different chaotic mixmaster collapse \cite{Peebles}.  
The universe approaches the singularity
 in a state that is wildly anisotropic, curved  and 
inhomogeneous -- a cosmological disaster.

However,  the story changes completely when we add an energy density component 
with 
$w>1$ \cite{chaos}.  The brane/scalar field kinetic energy density decreases as
\begin{equation}
\frac{\rho_{\phi}}{a^{3(1+w)}},
\end{equation}
where the exponent $3(1+w) >6$.  Now, the scalar field density grows faster than 
the 
anisotropy or any other terms as the universe contracts.  The longer the universe 
contracts, the more the scalar density dominates so that, by the bounce, the 
anisotropy 
and spatial curvature are completely negligible.  Also negligible are spatial 
gradients of 
fields.  The evolution is described by purely time-dependent factors, a situation 
referred 
to as {\it ultralocal}.  

In short, the striking discovery is that {\it a contracting universe with $w>1$ 
has the same 
effect in homogenizing, isotropizing and flattening the universe as an expanding 
universe 
with $w<-1/3$}.  This realization enables us to address the comment by some
 that the Cyclic Model is actually a model of inflation since it includes a period 
of dark 
energy domination that helps to homogenize and flatten the universe for the next 
cycle \cite{linde}.  
(Dark energy can be viewed as a form of ultra-slow inflation.)  

Now we see that this characterization is incorrect.  
As proof, consider an alternative model in which 
 the period of dark energy expansion is followed by a period of contraction with 
$w<1$.   
Despite being rather homogeneous and flat at the end of the dark energy dominated 
period, the universe would be highly inhomogeneous, anisotropic and randomly 
curved  
at the bounce.  The universe would, therefore,  
emerge in an unacceptable  state and would not satisfy the 
conditions for cycling.  

In short, the 
key element for making the universe homogeneous, 
isotropic and flat, for avoiding mixmaster behavior, and for insuring a cyclic 
evolution  is 
the contraction phase with $w>1$.

\subsection{$w>1$ and scale-invariant perturbations}

The fact that $w$ is greater than one
during the contraction phase is not only critical for the background 
homogeneous solution, but also for the perturbations \cite{duality,dual2}.  
Both the Cyclic Model and 
inflation generate density perturbations  from sub-horizon scale quantum 
fluctuations.  In each picture,
there is one phase when the sub-horizon scale fluctuations exit the horizon 
and a much later phase when they re-enter.  

In both cases, fluctuations leave the horizon because of the value of $w$, or, 
equivalently, $\epsilon \equiv \frac{3}{2} (1+ w)$.  For a universe with constant 
$\epsilon$, the scale factor $a(t)$ and the Hubble radius $H^{-1}$ are related by 
the 
Friedmann equations
\begin{equation}  \label{fried}
a(t) \sim t^{1/\epsilon} \sim (H^{-1})^{1/\epsilon}.
\end{equation}

If the universe is expanding and quantum fluctuations are supposed to leave the 
horizon, 
then it is necessary that $a$ grows faster than $H^{-1}$.  From the relation 
above,  
 this requires $\epsilon <1$ (or $w <-1/3$).  
 To obtain a nearly scale-invariant 
spectrum of fluctuations, it is necessary that $H^{-1}$ change very little during 
a period 
long enough for many modes to be stretched beyond the horizon.  This occurs in the 
limit 
$\epsilon \ll 1$.  In fact, the scalar spectral index $n_s$ about some given 
wavenumber 
can be computed by standard methods to be \cite{duality}
\begin{equation} \label{n1}
n_s -1 = -2 \epsilon + \frac{ d \, ln \, \epsilon}{dN}
\end{equation}
where  $N$ is a time-like variable that 
measures  the number of e-folds of inflation 
remaining when a given mode exits the horizon.   Here we see that, indeed, if 
$\epsilon$ 
is small and nearly constant, a nearly scale-invariant spectrum is predicted.

Now consider a contracting universe.  Both the wavelength of the fluctuations and 
the 
Hubble horizon are shrinking.  In order for a mode to exit the horizon, the 
horizon must 
shrink faster than the wavelength of the mode, or, equivalently, $H^{-1}$ must 
decreases 
{\it more rapidly} than $a(t)$.  According to (\ref{fried}), this requires 
$\epsilon>1$ or 
$w >-1/3$.  To obtain a spectrum that is nearly scale-invariant, we need $a$ to be 
nearly 
constant over a period when $H^{-1}$ changes a lot.  This occurs if $\epsilon \gg 
1$ or 
$w\gg 1$.  This is precisely what is obtained during the contraction phase of the 
Cyclic 
Model if the interbrane potential is negative and exponentially steep.  Using 
standard 
methods, we obtain for the spectral index \cite{duality}
\begin{equation} \label{n2}
n_s -1 = -\frac{2 }{\epsilon } - \frac{ d \, ln \, \epsilon}{dN}
\end{equation}
which, consistent with our intuitive analysis, predicts a nearly scale-invariant 
spectrum 
for $\epsilon \gg 1$ and nearly constant.

Comparing (\ref{n1}) and (\ref{n2}), we see that the two expressions are related 
by the 
transformation $\epsilon \rightarrow 1/\epsilon$.  That is, the perturbations 
produced in 
the contracting phase are dual to the perturbations in the expanding phase.  
Recently, 
L. Boyle {\it et al.} \cite{dual2} have shown that this surprising duality extends 
for all $\epsilon$.  
That is, {\it  for each inflating model with a scalar spectral index $n_s$, there 
is a 
corresponding contracting (ekpyrotic) model with the same $n_s$.)  }

A corollary is that the Cyclic model and inflation cannot be distinguished by 
observing 
the (linear) scalar perturbations alone.   We must dig further to determine if the 
perturbations were produced in a rapidly expanding phase or a slowly contracting 
phase.

One approach is to measure the spectral index of tensor perturbations 
\cite{cyclic,boyle}.  Here the two 
models make starkly different predictions. Inflation predicts a nearly scale-
invariant 
spectrum ($n_s \approx 0$) and the Cyclic Model predicts a very blue spectrum $n_s 
\approx 2$.  The difference arises because gravity plays a different role in the 
two 
scenarios. 

 In inflation, gravity plays a dominant role in expanding the universe and in 
creating a 
nearly de Sitter background that excites all light degrees of freedom.  Since the 
inflaton is 
nearly massless and the tensor modes are precisely massless and since they both 
evolve in the 
same (nearly) de Sitter background, both obtain a (nearly)
scale-invariant spectrum.  Furthermore, 
unless some 
additional parameters, fields and/or tunings
are introduced in the inflation model, the amplitude of the 
spectra are comparable.  

In the Cyclic Model, the situation is entirely different.  The universe is so 
slowly 
contracting that gravity is nearly irrelevant.  That is, the gravitational 
background is 
nearly Minkowski.  The scale-invariant fluctuations arise because one field, 
$\phi$, has a 
very steep potential.  As $\phi$ rolls down a steep potential, quantum 
fluctuations are 
unstable and amplified.  Since $\phi$ determines  the distance between branes, 
this means 
that the fluctuations in the time of collision grow.  For a negative, exponentially 
decreasing 
potential, one can show that $w \gg 1$ and, hence, the spectrum of $\phi$ fluctuations is nearly scale-invariant
(see discussion in Sec.~\ref{sec31}).
However, only 
$\phi$ obtains scale-invariant fluctuations since it is the only field with the 
steep 
potential.  All other light fields, including the tensor modes, only see a slowly 
contracting 
(nearly Minkowski) gravitational background.  This  background produces a blue 
spectrum.

The cosmological background in the Cyclic Model is not only slowly contracting, 
but the
value of the
Hubble parameter is exponentially small compared to inflation.  As a result, 
higher order non-gaussian corrections to the density perturbation spectrum, which 
are 
proportional to $H/M_{pl}$, are exponentially suppressed compared to inflation.  
Also, 
since $\phi$ is the only field to obtain scale-invariant fluctuations, it is not 
possible, for 
all practical purposes,
 to construct examples where perturbations in other degrees of freedom compete in 
producing the observed fluctuation spectrum.  This is why the perturbation 
spectrum 
predicted by the Cyclic Model is 
{\it super-gaussian} and {\it super-adiabatic} compared to 
inflation.  

Finally, we recall that there is an uncertain contribution to the inflationary 
prediction of density perturbations due to trans-Planckian effects \cite{trans}.  That is, the modes observed on the horizon today have wavelengths at the beginning of inflation that are smaller than the Planck scale 
(in most inflationary models).  The conventional calculations 
of the inflationary density perturbation spectrum
assume that the intial fluctuations on small scales corresponds to ground state fluctuations in empty, flat Minkowski space.  
This approximation may be invalidated for sub-Planckian fluctuations due to quantum gravity effects.  The magnitude of the correction is debatable.    
It is worth noting that sub-Planckian effects should not be a factor in 
cyclic models.  
This is because 
the perturbation
modes do not escape the horizon by having their wavelengths stretched.  
Rather, their wavelengths remain nearly constant while the horizon shrinks.   
The wavelengths of fluctuations observed in the cosmic microwave background
escaped the horizon during contracting when the Hubble radius was
exponentially larger than the Planck scale, and so sub-Planckian effects 
should be irrelevant.  Hence, the trans-Planckian issue is avoided in 
cyclic models and the prediction of density perturbations is more 
certain than inflation, where the introduction of additional fields 
and trans-Planckian physics can change nearly all of the predictions
of the simplest inflationary models.

\subsection{$w>1$ and the bounce}

Because $w>1$ during the contraction phase, the conditions leading up to the 
bounce are 
ultralocal \cite{chaos}.  Ultralocality means that the evolution equations are, to 
an
excellent 
approximation, only dependent on time-varying quantities. 
Spatial variations can be 
ignored.  One way to view the situation is that the Hubble horizon is contracting 
around 
any observer in the contracting spacetime, so differences between distant
points are 
irrelevant to the local evolution.  There is simply not time before the bounce for 
distant 
regions to interact.  Hence, instabilities are suppressed and evolution is smooth.
This simplicity is essential for analyzing
 the propagation of perturbations through the 
bounce.

The subject has been controversial in large part due to a subtle
difference in the nature of 
 growing and decaying modes  in a contracting 
phase compared to an expanding phase.  In inflation, the curvature fluctuation on 
comoving hypersurfaces is a familiar growing mode, and the time delay is a 
decaying 
mode; but, the roles are reversed in a contracting phase.  Hence, the curvature 
fluctuation 
on comoving hypersurfaces shrinks to zero as the bounce approaches. 

One of the theorems learned from studying perturbations in
inflation is that the curvature 
fluctuation is conserved for modes outside the horizon \cite{Bardeen}.  If this 
remained true for the 
Cyclic Model, then zero curvature fluctuation before the bounce implies zero 
curvature 
fluctuation after the bounce. A corollary
is that the growing time-delay mode during contraction 
transforms entirely 
into a decaying mode after the bounce, and there is  no scale-invariant 
spectrum of curvature perturbations in the expanding phase.  
This would be correct if the bounce occurs on a comoving 
hypersurface so that the curvature perturbation can be matched continuously.  

However,
in four dimensions, the bounce is singular in all dimensions, and the choice of 
matching 
hypersurface at the bounce is unclear.  
Some have nevertheless argued that a comoving 
hypersurface is the only plausible choice \cite{contrascale}.  They conclude that 
no curvature fluctuations 
exist after the bounce and the model fails.  Others choose a different 
hypersurface for the 
bounce, in which case the curvature fluctuation is not continuous at the bounce 
\cite{ekperts}.  
For any choice of matching surface {\it other than} the comoving hypersurface,
the growing, time-delay mode during contraction transforms into a mixture of 
decaying 
(time-delay) mode and growing (curvature fluctuation)
mode after the bounce \cite{durrer}.  The 
curvature fluctuation, thereby, inherits a fraction of the initial time-delay mode 
and 
produces an observationally acceptable,  nearly-scale invariant spectrum of 
fluctuations 
in the expanding phase.     Others claim that the matching hypersurface is 
ambiguous \cite{durrer,ambig}, and 
so the Cyclic Model makes no definite prediction for the density perturbation 
amplitude. 

In the Cyclic Model, though, the four-dimensional picture is only an approximation 
describing a higher dimensional picture of colliding branes.  The branes make the 
critical 
difference \cite{newtolley,craps,newbrand}.  The branes
exist before, during and after the bounce, and they fix a precise 
hypersurface for the bounce.  Namely, the bounce corresponds to a time-slice in 
which 
each point on one brane is in contact with a point on the other brane.  That is, 
the 
appropriate  matching hypersurface
is the one  in which the bounce is everywhere simultaneous \cite{newtolley,nobel}.  
Since 
$\phi$ is the modulus field that determines the bounce between branes, one might 
imagine 
that this corresponds to $\delta \phi=0$, which is a comoving hypersurface.  
However, 
$\phi$ only measures the distance between branes if they are static. If the branes 
are 
moving, there are corrections to the distance relation that depend on the brane 
speed and 
the bulk curvature scale \cite{newtolley}.  Roughly speaking, the  condition that 
$\delta \phi=0$ ensures no 
velocity perturbations tangent to the branes, but simultaneous bounce requires no 
relative 
velocity perturbations perpendicular to the branes.  The gauge transformation 
required to 
transform from $\delta \phi =0$ to the proper gauge introduces a scale-invariant 
curvature 
perturbation on the surface of collision.  Thus, in a collision-simultaneous 
gauge, the 
spatial metrics on the branes {\it necessarily} acquire long wavelength, nearly 
scale 
invariant curvature perturbations at the collision.   This result has been 
obtained using 
several different methods by  three independent groups 
\cite{newtolley,craps,newbrand}.

The result for the long wavelength curvature perturbation  
amplitude in the four-dimensional effective theory,  
propagated into the hot big bang after the brane collision  
is \cite{newtolley,design}: 
\begin{equation}  \label{zeta}
\zeta_M = { 9 \epsilon_0 \over 16 k^2 L^2} {\tanh (\theta/2) \over 
\cosh^2 (\theta/2)} (\theta-\sinh \theta) \approx 
- { 3 \epsilon_0 V_{coll}^4 \over 64 k^2 L^2}  
\end{equation} 
where $\theta$ is the rapidity corresponding to the relative 
speed $V_{coll}$ of the branes at collision, and the second formula 
assumes $V_{coll}$ is small. $L$ is the bulk curvature 
scale, and $ \epsilon_0 \over 16 k^2$ has  
a scale invariant power spectrum. The presence of radiation on 
the branes before or after the collision produces an  
additional correction term given in full in 
Ref.~\ref{newtolley}. 
As discussed above, the physical origin of the curvature 
perturbation is in the time delay between the collision 
timeslice and the $\delta \varphi=0$ (or comoving) hypersurfaces, on 
which 
the branes do not possess long wavelength scale invariant 
curvature perturbations. 

We emphasize that the key to this  result is the 
branes -- a new, unanticipated  aspect of branes 
that turns out to be  essential to the Cyclic Model.
The orbifolding of the extra dimensions means that 
the extra dimension is non-uniform. 
In particular, as the branes approach, we 
found that bulk 
excitations perturb the collision-simultaneous time-slice
from comoving.  For comparison, if we consider
an 
extra dimension that is Kaluza-Klein compactified, we 
reach a different conclusion.  Then, the comoving slicing is well-defined up
to and including the bounce and no scale-invariant
perturbations propagate into the expanding phase.

\section{Onward to the non-linear regime}

Thus far, we have analyzed the propagation of perturbations 
through the bounce assuming  they are linear.  
From this, we have learned that branes introduce a new physical
element essential for propagating perturbations through the bounce.
Also, we have obtained, we hope, a good estimate of the spectral
amplitude.  However, a full analysis including the non-linear 
physics close to the bounce is required to complete the picture.

 Because $w>1$ during the contraction phase, the universe remains 
nearly homogeneous, isotropic and flat and the linear approximation 
remains valid up until the branes are about a string scale-length 
apart.  At this point, corrections to the Einstein action become 
important.  We cannot be sure what those corrections are.
However, assuming there is a bounce,
 they operate for a very short time.   The 
branes lie within a string-scale-length for 
roughly a string scale-time, or about $10^{-40}$~seconds. 

During these last instants before the bounce, the modes of interest for 
cosmology, {\it e.g.,} wavelengths which lead to the formation of 
galaxies and larger scale structure, are far outside the horizon and 
their dynamics is frozen.  Although their 
amplitude may become non-linear, there is not enough time 
during the bounce for interactions to 
alter the long-range correlations.   Hence, we conjecture, it is 
reasonable to match the linear behavior just before the bounce to the 
linear behavior just after the bounce.

In fact,
one approach to the matching problem may be  to avoid $t=0$  altogether
by analytically continuing in the complex $t$-plane 
in a semicircle with radius greater than
the string scale and  connecting
 negative to positive real values of $t$ \cite{nobel}.
Then, the linear analysis described   above
would remain essentially uncorrected by nonlinear gravitational
effects, at least on long (three-dimensional) wavelengths.
Work is currently in progress to construct such
a continuation in nonlinear gravity.

On the other hand, for modes with wavelength less than $10^{-30}$ cm, 
causal dynamics can alter the dynamics in the final instants.  In 
particular, the non-linear corrections to gravity could conceivably 
produce  large amplitude effects that lead to the formation of many 
tiny primordial black holes.  

The black holes are bad news for those wishing to track 
precisely what occurs at the bounce.  String theoretic methods are 
probably not powerful enough 
to analyze precisely this kind of inhomogeneous, 
non-linear regime.  However, from a cosmological point-of-view, it is 
straightforward to envisage their effect, assuming that the branes 
bounce.  

The black holes are small and have a mass roughly of order the mass 
density times Hubble volume at the collision.  This scale is 
model-dependent, but for the wide range of 
parameters allowed for the cyclic 
model based on other constraints \cite{design}, the mass is 
sufficiently small that the black holes decay in much less than one 
second, well before primordial nucleosynthesis. 

We conjecture that the black holes can be a boon to the scenario.  
(See also \cite{banks}, who consider an alternative model that begins with a
dense gas of black holes.) 
Their lifetime is long enough that they likely dominate the energy 
density before they decay.  Consequently, their evaporation provides 
the entropy observed today.  When they decay, their temperature rises 
near the end to values high enough to produce massive particles with 
baryon-number violating decays.  At this point, the black holes are 
much hotter than the average temperature of the universe, so the decays 
occur when the universe is far from equilibrium.  Assuming CP-violating 
interactions also exist, as in conventional high-temperature 
baryogenesis scenarios, the decay can produce the observed baryon 
asymmetry \cite{baumann}.  In addition, the decay can produce dark 
matter particles that can meet current observational constraints.

\section{What we have learned}

Our study of the Cyclic Model has uncovered  surprising new facts about 
contracting universes.  Namely,  a contracting universe with $w>1$ has 
remarkable properties analogous to an expanding phase with $w \le -
1/3$.  The homogeneous solution to the Friedmann equation becomes 
spatially uniform, isotropic and flat.  At the perturbative level, a 
nearly scale-invariant spectrum of density perturbations is generated.  
There is a precise duality relating (linear) scalar perturbations 
produced in an inflating phase to those produced in a contracting
(ekpyrotic) phase.
The evolution equations are also ultralocal (purely time-dependent) at 
least up until stringy corrections to the Einstein equation become 
significant when the branes are separated by less than a string scale-
length.  

Consequently, many of the conventional worries of the past about 
contracting phases are addressed, and attention is turning to what 
happens in the final instants before and after the collision.  The goal 
is to determine if: (a) the bounce occurs; and, (b) perturbations on 
wavelengths greater than the string scale lengths (which includes the 
wavelengths of cosmological interest) obey the matching rule naturally 
inferred by analytically continuing the linear solution.   Current
research is focused on these exploring these issues.

The outcome has profound implications for cosmology and fundamental
physics. If inflation is correct,
then we are blocked from any direct knowledge of the big bang and any other 
pre-inflationary conditions by a period of superluminal expansion. If the Cyclic Model is correct, then our
measurements of  microwave background fluctuations and large-scale structure
are 
{\it to leading order} direct probes 
of the big crunch and big bang, including
stringy and extra-dimensional physics, as illustrated by Eq.~(\ref{zeta}).  
Settling the cosmological issue of whether the density
fluctuations were produced during a period of expansion or contraction (by
searching for tensor fluctuations, non-gaussianity and non-adiabaticity) will 
also determine whether
physical conditions near the big bang can be probed empirically or not.  This
raises the stakes and enhances the importance of distinguishing the two
scenarios.

\newpage
{\large \bf Acknowledgments} We thank  our collaborators,
L. Boyle, J. Erickson, S. Gratton, J. Khoury, A. Tolley, and D. Wesley,
for their important contributions to the work discussed herein, and,
also, M. Kleban, R. Bousso, E. Flanagan,  and L. Susskind
for  valuable discussions. 
The work of NT was partially supported by PPARC (UK), and
that of PJS by US Department of Energy Grant DE-FG02-91ER40671 (PJS).
PJS is also Keck Distinguished Visiting Professor at
the Institute for Advanced Study with support from the Wm.~Keck
Foundation and the Monell Foundation.

\end{document}